\documentclass[conference,10pt]{IEEEtran}
\usepackage{graphicx,epstopdf,amsmath,epsfig,times,setspace,bm}
\IEEEoverridecommandlockouts
\usepackage{bm}
\usepackage{array}
\usepackage{subfig}

\usepackage[T1]{fontenc}
\usepackage{times}
\usepackage[mathscr]{eucal}
\usepackage{mathptmx}

\usepackage{multirow}

\let\mathcal\relax
\DeclareMathAlphabet\mathcal{OMS} {cmsy}{b}{n}
\SetMathAlphabet \mathcal{normal}{OMS}{cmsy}{m}{n}
\DeclareMathAlphabet\mathbcal{OMS} {cmsy}{b}{n}

\DeclareMathOperator*{\argmin}{\arg\!\min}

\newcommand{\cE}{\mathcal{E}}

\newcommand{\cP}{\mathcal{P}}

\newcommand{\equals}{=}
\newcommand{\minus}{\!-\!}
\newcommand{\gteq}{\!\ge\!}

\newcommand{\plus}{\!+\!}

\newcommand{\dB}{\, \mathrm{dB}}

\begin{document}
\title{Allocation of Repetition Redundancy in LoRa \vspace*{-12mm}\\
{\footnotesize \textsuperscript{}}
\thanks{This work has been supported by the K-project DeSSnet (Dependable, secure and time-aware sensor networks), which is funded within the context of COMET -- Competence Centers for Excellent Technologies by the Austrian Ministry for Transport, Innovation and Technology (BMVIT), the Federal Ministry for Digital and Economic Affairs (BMDW), and the federal states of Styria and Carinthia; the COMET program is conducted by the Austrian Research Promotion Agency~(FFG).}
}

\author{\IEEEauthorblockN{Siddhartha Borkotoky\textsuperscript{1}, Christian Bettstetter\textsuperscript{1,2}, Udo Schilcher\textsuperscript{1}, and Christian Raffelsberger\textsuperscript{1}}
\IEEEauthorblockA{\textit{\textsuperscript{1}Lakeside Labs GmbH, \textsuperscript{2}Institute of Networked and Embedded
Systems, University of Klagenfurt} \\
9020 Klagenfurt am W\"{o}rthersee, Austria \\
borkotoky@lakeside-labs.com}
}

\maketitle

\begin{abstract}
We consider a multipoint-to-point network in which sensors periodically send measurements to a gateway. The system uses Long Range (LoRa) communications in a frequency band with duty-cycle limits. Our aim is to enhance the reliability of the measurement transmissions.

In this setting, retransmission protocols do not scale well with the number of sensors as the duty cycle limit prevents a gateway from acknowledging all receptions if there are many sensors. We thus intend to improve the reliability without acknowledgments by transmitting multiple copies of a measurement, so that the gateway is able to obtain this measurement as long as it receives at least one copy. Each frame includes the current and a few past measurements.

We propose a strategy for redundancy allocation that takes into account the effects of fading and interference to determine the number of measurements to be included in a frame. Numerical results obtained using the simulation tool LoRaSim show that the allocation of redundancy provides up to six orders of magnitude decrease in the outage probability. Compared to a system that blindly allocates the maximum redundancy possible under duty-cycle and delay constraints of the gateway and memory constraints of the sensors, our technique provides up to $30\,\%$ reduction in the average energy spent to successfully deliver a measurement to the gateway.
\end{abstract}



\renewcommand{\baselinestretch}{0.9}
\small\normalsize

\section{Introduction}
\label{intro}
Low power wide area (LPWA) communication technologies~\cite{RKS17} are potential enablers for a variety of Internet of Things applications. Their salient features are long transmission ranges and low power consumption achieved at the expense of low data rates. They are attractive choices for applications that do not have stringent delay and  throughput requirements but must support many devices distributed over a large area at low operational cost. Applications can be found in the domains of smart city, smart metering, industrial monitoring, and agriculture, to give some examples.

Examples of LPWA technologies are LoRa, Sigfox, Ingenu, and Weightless~\cite{RKS17}. Our focus here is on LoRa, which has received significant attention over the past few years. LoRa defines a spread-spectrum physical layer that exhibits high robustness against thermal noise and multipath effects. LoRaWAN defines a medium access control (MAC) layer built on top of its physical layer. End devices transmit LoRa frames to a gateway, which forwards the frames to the desired destination, typically over a wired connection. The duty cycle of LoRa devices operating in the 868 MHz frequency band in the European Union must not exceed 1\%~\cite{AVT17}.

Despite their robustness, LoRa transmissions can still be disturbed by temporal variations in the radio links, such as fading and shadowing, which result in the loss of frames at the gateway~\cite{MRP17}. Another reason for frame loss is interference from other LoRa transmitters~\cite{BRV16}. Automatic Repeat Request or similar techniques that rely on acknowledgments and retransmissions do not provide scalable solutions for LoRa networks because the number of transmissions that can be made by a node is restricted by the limits on the duty cycle. LoRa devices of a certain class listen for only a short period of time following a transmission; the gateway's acknowledgement is lost unless it is sent within that period. These limitations make it infeasible for a gateway to acknowledge receptions when the number of end devices is large. Approaches that do not rely on acknowledgments from the gateway are therefore of practical interest.

\begin{figure}
\centering
\includegraphics[scale=0.25,bb=320 70 500 540]{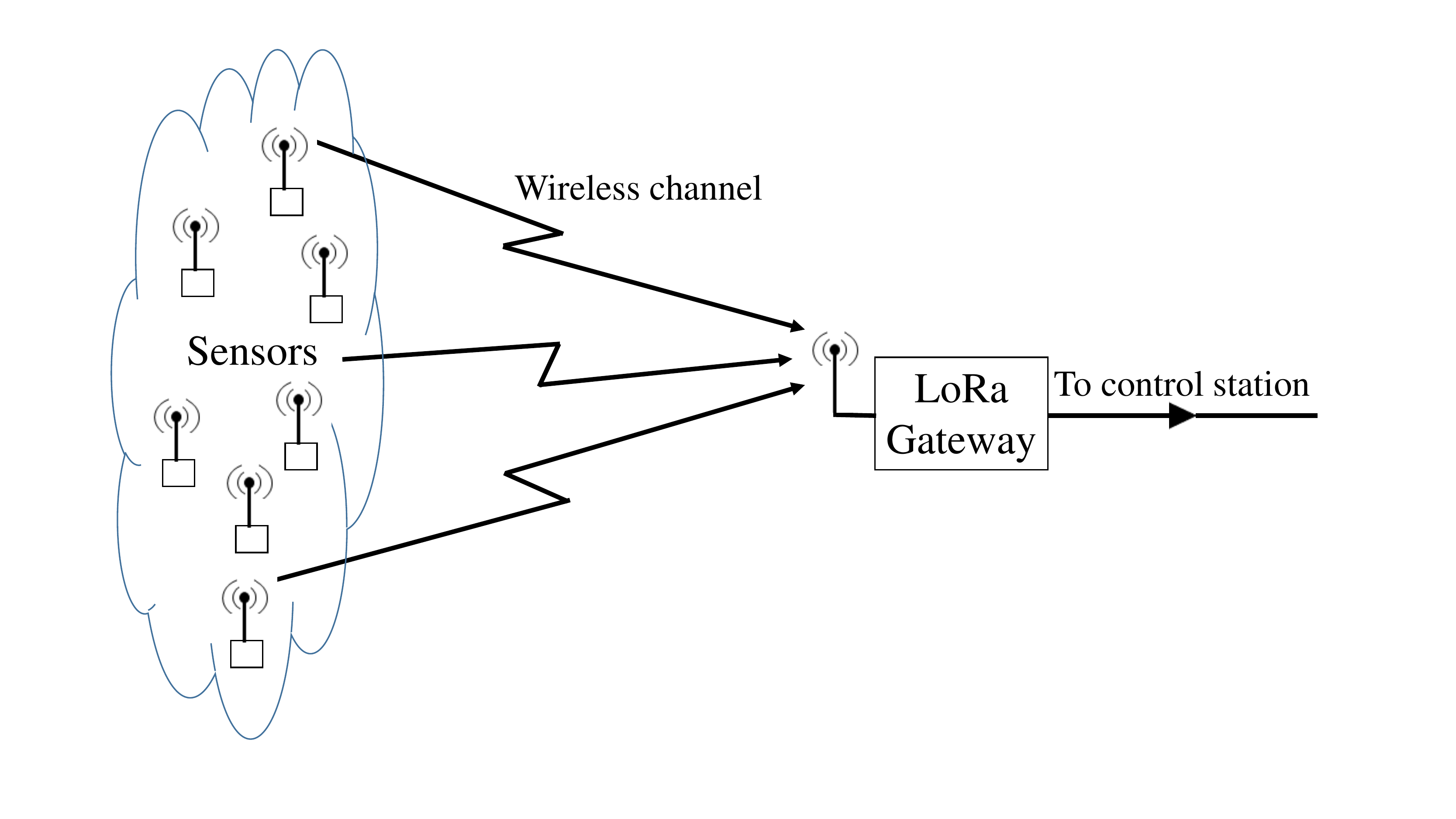}
\setlength{\belowcaptionskip}{-18pt}
\caption{LoRa sensor network.}
\label{net_diagram}
\end{figure}
\renewcommand{\baselinestretch}{0.9}
\small\normalsize

Here, we consider a wireless sensor network in which the sensors use LoRa to communicate with a gateway (see Fig.~\ref{net_diagram}). There is no acknowledgment from the gateway in response to received frames. Instead, to improve the probability of successful delivery, sensors transmit redundant information by including the current and a few past measurements in each frame. A measurement reaches the gateway as long as at least one of the frames that carries it is delivered.

The number of redundant transmissions of a measurement is crucial to the performance. On one hand, sending too few redundant copies may not sufficiently mitigate the effects of frame loss. On the other hand, the duration of a frame increases with increased redundancy, which in turn increases the energy expenditure and probability of collisions. The amount of redundancy is also constrained by the time after which a measurement is no longer of interest to the system, the maximum number of past measurements a sensor can store in its memory, and the maximum duty cycle. We provide a method to determine the amount of redundancy that must be included per LoRa frame to achieve a given probability of successful delivery of the sensor data to the gateway. This is achieved by expressing the probability as a function of the redundancy, and then evaluating the expressions to find the redundancy that provides the target performance. The derivation of the expressions takes into account the effects of interference and fading. The expressions are evaluated subject to the duty-cycle limit, the memory constraints of the sensors, and the maximum tolerable delay of the sensing application. The evaluations are performed at the gateway side, so that the sensors are not computationally burdened. We carry out simulations in LoRaSim~\cite{LoRaSim} to demonstrate the benefits of our approach.

\section{Related Work}
The transmission of redundant data for improved reliability of LoRaWAN is also considered in~\cite{MRP17}, where the redundancy is generated by performing application-layer coding on past data. To avoid increasing the computational load at the sensors, we do not consider coding. Unlike our work, the presence of multiple transmitters and the resulting interference are not considered in~\cite{MRP17}.

There is a wide range of results available on interference analysis based on tools from stochastic geometry~\cite{SBB12,CTS15}. They consider the sum interference power as opposed to the power of the strongest interferer. Since only the latter is important for LoRa due to the capture effect~\cite{GeR17}, these results are not applicable for our purpose. Results on the outage probability of LoRa that take into account the capture effect are given in~\cite{GeR17}. The scenario analyzed in~\cite{GeR17} comprises a random number of nodes distributed over a circular region with the gateway at the center. In contrast, our work is motivated mainly by use cases pertaining to industrial monitoring in which a known number of sensors are placed over a small area. The area could be a production floor, a processing unit, a storage facility, or a collection of entities equipped with many sensors to periodically measure physical quantities of interest. The gateway could be located at an office building and wired to a computational unit where the sensor data are processed and monitored.

\section{Characteristics of LoRa} \label{LoRa_description}
The LoRa physical layer employs chirp spread spectrum to modulate the information bits~\cite{Sem15}. The LoRa waveform is characterized by its \emph{spreading factor}, which determines the ratio between the symbol rate and the chip rate and can take integer values between 7 and 12. The duration of a symbol in a LoRa frame that employs a spreading factor~$s$ and has bandwidth $w$ is given by
\begin{eqnarray}
t_{\mathrm{sym}} &=& 2^{s}/w\:.
\end{eqnarray}
A higher spreading factor provides greater immunity to thermal noise but also increases the frame duration, which doubles as the spreading factor is increased by one. For additional protection against channel disturbances, a channel code of rate 4/5, 4/6, 4/7, or 4/8 is employed~\cite{Sem13}.  The bandwidth of the LoRa waveform can range from 7.8~kHz to 500 kHz, although typically employed values are 125~kHz, 250 kHz, and 500 kHz~\cite{BRV16}. The center frequency is programmable between 137 MHz to 1020 MHz in steps of 61 Hz~\cite{BRV16}. A transmitter can perform channel hopping by changing the center frequency in a pseudo-random manner from frame to frame.

The payload in a frame is preceded by a preamble for synchronization and an optional header. The duration of a frame that contains $b$ bytes of payload is
\begin{eqnarray} \label{frame_duration}
t_{\mathrm{fr}}(b) &=& t_{\mathrm{pr}} + t_{\mathrm{pl}}(b)\:,
\end{eqnarray}
where $t_{\mathrm{pr}}$ and $t_{\mathrm{pl}}(b)$ are the durations of the preamble and payload portions, respectively. The preamble duration is
\begin{eqnarray}
t_{\mathrm{pr}} &=& (n_{\mathrm{pr}} + 4.25) \, t_{\mathrm{sym}}\:,
\end{eqnarray}
where $n_{\mathrm{pr}}$ is the number of preamble symbols in the frame~\cite{Sem13}. The payload duration is
\begin{align} \label{payload_duration}
t_{\mathrm{pl}}(b) &= \left[8 + \max\left\{ \left\lceil \frac{2b-s-5h+11}{s-2l} \right\rceil (c+4), 0\right\}\right]\,t_{\mathrm{sym}}\:,
\end{align}
where $h$ is 1 when a header is included and 0 otherwise, $l$ is 1 when low data rate optimization is enabled and 0 otherwise, and $c$ depends on the rate of the channel code employed for the frame and can take integer values between 1 and 4~\cite{Sem13}.

A LoRaWAN has a star topology, where the end devices communicate with one or more gateways. A gateway forwards the frames received from end devices to a network server, which in turn connects to an application server. Three classes of operation are defined for end devices~\cite{AVT17}: In Class A, an end device employs unslotted ALOHA for uplink transmissions, and after a transmission, the device listens during two receive windows for any response from the gateway. Class B allows for additional receive windows, which are scheduled using synchronizing beacons from the gateway. In Class C, the end devices are always in the receive mode unless they are transmitting. All end devices must support Class A, whereas the other classes are optional. In this paper, we consider networks in which multiple end devices operate in Class A to transmit data to a single gateway operating in Class C. All devices are subject to a duty-cycle limit of $1\,\%$.

LoRa waveforms that employ different spreading factors are orthogonal to one another. Therefore, only those frames that employ the same spreading factor and overlap in frequency can interfere. Because chirp spread spectrum is a form of frequency modulation, LoRa waveforms exhibit the capture effect: Suppose two signals having the same spreading factor and using the same channel arrive simultaneously at the receiver. If one signal is stronger than the other, the strong signal is correctly demodulated and decoded, whereas the frame carried by the weaker signal is lost. Field measurements show that a LoRa frame is correctly received if the strongest of the interfering signals is at least $6\dB$ weaker than the desired signal.

\section{Analysis of Failure Probability} \label{reliability_analysis}
Consider $n$ sensors sensing physical parameters every $t$ seconds and transmitting their measurements to a gateway using Class~A channel access. All sensors have the same transmission power and are not time synchronized. A frame contains the current measurement and $r$ past measurements. In the following, we derive the \emph{failure probability}, defined as the probability that the gateway fails to obtain a measurement, as a function of $r$.

A frame is delivered to the gateway if neither interference nor fading results in the loss of the frame. As mentioned earlier, only the frames that employ the same spreading factor can interfere with one another. Furthermore, frame loss due to fading is independent of the presence of other transmitters. Therefore, it suffices to consider only those sensors that transmit with the same spreading factor. Without loss of generality, we assume that the same spreading factor is employed by all $n$ sensors.

The duty cycle $f(r)$ of a sensor, defined as the fraction of time the sensor is in the transmit mode, is
\begin{eqnarray}
f(r) &=& \frac{t_{\mathrm{fr}}((r+1)\beta)}{t}\:,
\end{eqnarray}
where $\beta$ is the number of bytes required to represent one sensor measurement.

The power received by the gateway from sensor $i$ is
\begin{eqnarray} \nonumber
P_i &=& A_i\left( \frac{\lambda}{4\pi D_i} \right)^\alpha \cP \\
&=& \gamma A_i D_i^{-\alpha}\:,
\end{eqnarray}
where $A_i$ is the gain due to fading, $\lambda$ is the signal wavelength, $\alpha$ is the pathloss exponent, $D_i$ is the Euclidean distance from the gateway to the $i$th sensor, $\cP$ is the transmit power, and $\gamma = (\lambda/4\pi)^{\alpha}\cP$. We assume that the $A_i$ and $D_i$ are independent and identically distributed (i.i.d.). It follows that the $P_i$ are also i.i.d. Hence, we drop the subscript $i$ from these terms when dealing with their probability distributions in the remainder of the paper.

The cumulative distribution function (cdf) for the received power conditioned on the distance $D$ is
\begin{eqnarray} \nonumber
F_{P}(x | D = u) &=& P(\gamma A u^{-\alpha} \leq x) \\
&=& F_A(\gamma^{-1} u^\alpha x)\:,
\end{eqnarray}
where $F_A(\cdot)$ denotes the cdf for the gain $A$. We remove the conditioning to obtain
\begin{eqnarray}
F_{P}(x) &=& \int\displaylimits_{S_D} F_A(\gamma^{-1} u^\alpha x)\, f_D(u)\:\mathrm{d}u\:,
\end{eqnarray}
where $f_D(\cdot)$ is the probability distribution function for the distance $D$ and $S_D$ is the support of $f_D(\cdot)$.

Now we consider the reception of a frame at the gateway. The number of other nodes $N_t$ transmitting at the same time and frequency, causing interference, is a \mbox{binomial$(n \minus 1,qf(r))$} random variable, where $q$ is the probability of a transmitter choosing the same channel. If all nodes transmit on the same channel, then $q = 1$. If the transmitters hop among $u$ channels uniformly at random, then $q = 1/u$. We are interested in scenarios with moderate to large values of $n$ (e.g., 30 or greater). To satisfy the limits on the LoRa duty cycle, the value of $f(r)$ can be at most $0.01$. Due to these attributes of $n$ and $f(r)$, the probability mass function for $N_t$ can be approximated by the Poisson distribution with
\begin{eqnarray}
P(N_t = k) &\approx& \frac{v^k(r)e^{-v(r)}}{k!}\mbox{ for }k \geq 0\:,
\end{eqnarray}
where $v(r) = (n - 1) q f(r)$. Let $M$ denote the power of the signal from the strongest interferer. Because the $P_i$ are i.i.d., $M$ has the following conditional cdf:
\begin{eqnarray}
F_M(x | N_t = k) &=& F_P^k(x)\:.
\end{eqnarray}
Removing the conditioning, we obtain
\begin{eqnarray} \nonumber
F_{M}(x) &=& \sum_{k=0}^{n-1} F_M(x | N_t = k) P(N_t = k) \\ \nonumber
&\approx& \sum_{k=0}^{n-1} F_P^k(x) \frac{v^k(r)e^{-v(r)}}{k!} \\ \nonumber
&=& e^{-v(r)} \sum_{k=0}^{n-1} \frac{[v(r)F_P(x)]^k}{k!} \\ \nonumber
&\stackrel{(a)}{\approx}&  e^{-v(r)} e^{v(r)F_P(x)} \\ \nonumber
&=& e^{-v(r)[1-F_P(x)]} \\
&=& e^{-v(r)[1-\int\displaylimits_{S_D} F_A(\gamma^{-1} u^\alpha x) f_D(u)\:\mathrm{d}u]}\:.
\end{eqnarray}
In $(a)$, it is assumed that the values of the summands for \mbox{$k \gteq n$} are negligible, which is a reasonable assumption for the $n$-values of interest. For example, if \mbox{$F_p(x) \equals 1$}, \mbox{$q \equals 1$}, and $f(r) \equals 0.01$, which are the maximum possible values for each of these quantities, the summands are less than $10^{-79}$ for $n \geq 50$.

Conditioned on the distance $w$ between gateway and desired sender and on the fading coefficient $a$ of the desired signal, the probability of an outage due to interference is \mbox{$1 \minus  F_M (0.25 \gamma a w^{-\alpha})$}. The conditioning is removed to obtain
\begin{eqnarray} \nonumber \label{S_I_general}
P_i(r) &=& \int\displaylimits_{S_D} \!\int\displaylimits_{S_A} (1 - F_M (0.25 \gamma a w^{-\alpha})) f_A(a)f_D(w)\:\mathrm{d}a\:\mathrm{d}w \\
&=& 1 - \int\displaylimits_{S_D} \!\int\displaylimits_{S_A}  e^{-\kappa(a,w,r)} f_A(a)f_D(w)\:\mathrm{d}a\:\mathrm{d}w,
\end{eqnarray}
where \mbox{$\kappa(a,w,r) = v(r)[1- \int_{S_D} F_A(0.25 a u^{\alpha} w^{-\alpha}) f_D(u)\:\mathrm{d}u]$} and $S_A$ is the support of $f_A(\cdot)$.

For the special case in which the sensors form a dense cluster and the gateway is distant enough from this cluster such that the distances between the sensors are negligible relative to their distance from the gateway, it is possible to ignore the differences in path loss, yielding an approximation:
\begin{eqnarray} \label{S_I_equal_dist_assumption}
P_i(r) &\approx& 1 - \int\displaylimits_{S_A} e^{-v(r)[1- F_A(0.25 a)]} f_A(a)\:\mathrm{d}a\:.
\end{eqnarray}

The probability of an outage due to fading is the probability that the received power is below the receiver's sensitivity $s$, which is
\begin{eqnarray} \label{S_f_general} \nonumber
P_f(r) &=& \int\displaylimits_{S_D} P(\gamma A u^{-\alpha} < s)\, f_D(u) \:\mathrm{d}u\\
&=& \int\displaylimits_{S_D} F_A(\gamma^{-1} u^{\alpha} s)\,f_D(u)\:\mathrm{d}u\:.
\end{eqnarray}

If the distances between the sensors are negligible relative to their distance from the gateway, we have
\begin{eqnarray} \label{S_F_equal_dist_assumption}
P_f(r) &\approx& F_A(\gamma^{-1}d^{\alpha}s)\:,
\end{eqnarray}
where $d$ is the approximate distance between gateway and sensors.

The failure probability is the probability that the gateway does not receive any of the $r \plus 1$ frames containing a given measurement, which is
\begin{eqnarray} \label{combined_success_prob}
P_{\mathrm{fail}}(r) &=& (1 - (1-P_i(r))(1-P_f(r)))^{r+1}\:.
\end{eqnarray}

\section{Procedure for Redundancy Allocation} \label{optimal_r}
The system has a maximum tolerable delay $d_{\max}$, which is the time after which a measurement is no longer of interest to the system. Since a node transmits every $t$ seconds, a measurement can be delivered with a maximum delay of $rt$ seconds if $r$ past measurements are included per frame. Let $b_{\max}$ be the maximum number of measurements that can be stored in a sensor's memory. Further, let $\hat{r}_{\max}$ be the maximum number of redundant measurements that can be included in a frame without violating the duty-cycle limit; that is, $\hat{r}_{\max} = \max\{r : f(r) \leq 0.01\}$.

Then, the maximum number of past measurements that can be included in a frame is
\begin{eqnarray}
r_{\max} &=& \min \{d_{\max} / t, b_{\max}, \hat{r}_{\max}\}\:.
\end{eqnarray}

Suppose that we do not wish to exceed a failure probability $P_t$. In order to determine the amount of redundancy to use, the gateway first determines $r^*$ by
\begin{eqnarray}  \label{r_star}
r^* &=& \min \{r: P_{\mathrm{fail}}(r) \leq P_t, r \leq r_{\max}\}\:.
\end{eqnarray}
If no value of $r$ satisfies $P_{\mathrm{fail}}(r) \leq P_t$, $r^*$ is determined by
\begin{eqnarray}  \label{r_star_1}
r^* &=& \argmin_r \{P_{\mathrm{fail}}(r): r \leq r_{\max}\}\:.
\end{eqnarray}

An examination of \eqref{frame_duration}--\eqref{payload_duration} reveals that, for a given spreading factor, multiple payload sizes result in LoRa frames that have the same duration and hence the same transmission energy. This is a result of the ceiling operation in~\eqref{payload_duration}. For example, for a spreading factor of 10, payloads of 1 byte through 4 bytes all produce frames of the same duration.
Hence, if there exists an $\tilde r > r^*$, such that both $r^*$ and $\tilde r$ yield the same frame length, then by using $\tilde r$ instead of $r^*$, one could achieve superior performance without spending additional energy. Therefore, the gateway computes the number of most recent past measurements to be included per frame:
\begin{eqnarray}  \label{r_tilde}
\tilde r = \max \{r : t_{\mathrm{fr}}((r + 1)\beta) = t_{\mathrm{fr}}((r^* + 1)\beta), r \leq r_{\max}\}\:.
\end{eqnarray}

If not all end devices use the same spreading factor, the described procedure is carried out for each spreading factor individually.
Due to their orthogonality, the redundancy allocation is independent for different spreading factors.

\section{Simulation Setup}
\label{sim_setup}

LoRaSim is used to simulate the transmission of measurements from $n$ sensors to a gateway. We modify the software to implement independent Nakagami-$m$ fading on the links and periodic transmissions (in contrast to LoRaSim's default option of exponentially distributed transmission intervals). The gateway and sensors are simulated as points on a two-dimensional plane, with the gateway at the origin. The coordinates of sensor $i$ are $(X_i, Y_i)$, where $X_i$ and $Y_i$ are uniform random variables in the range [30 m, 42 m]. It follows that the minimum and maximum possible distances between the sensors and the gateway are 42.4 m and 59.4 m, respectively. Each sensor transmits a measurement of size 1 byte every $t = 30$ seconds. The sensors can store at most $b_{\max} = 10$ bytes of data, and the maximum tolerable delay for the network is $d_{\max} \equals 4.5$ minutes. The links from the sensors to the gateway experience Nakagami-$m$ fading with $m \equals 1$, which is equivalent to Rayleigh fading. The pathloss exponent is $\alpha = 4$. The frames employ a spreading factor of 10. For this spreading factor, the power of the received signal from a sensor in the absence of fading is above the sensitivity of the gateway's radio receiver. However, fading and interference may cause frame outages. Each transmission occupies a bandwidth of 125 kHz. We employ LoRaSim's native setting for channel hopping and transmission power: A sender chooses one of three center frequencies (860 MHz, 864 MHz, and 868 MHz) uniformly at random for each frame and uses a transmit power of 14\:dBm. A channel code of rate 4/5 is used. From the values of $d_{\max}$, $b_{\max}$, and the choice of the spreading factor, we can determine that $r_{\max} = 9$.  The simulation setup is intended to mimic certain industrial use cases currently under investigation by the authors. These involve a chemical production plant, in which sensors are installed for the purpose of predictive maintenance. In such a setup, many sensors could be installed over a small region, such as a room. The gateway is located at a control station in a different room; there is no line-of-sight path between the sensors and the gateway, and the pathloss is relatively high.

Two metrics are employed for performance evaluations: The \emph{measurement loss rate} (MLR) is the fraction of sensor measurements that the gateway fails to receive. The \emph{energy expenditure per delivered measurement} ($E_m$) is the average transmission energy in millijoules that a sensor spends in order to successfully deliver a measurement to the gateway; it is obtained by dividing the energy required to transmit one LoRa frame by $1 - \mathrm{MLR}$.

For each simulation run, we simulate three hours of network operation. The output of the $i$th run is the fraction of frames, $\rho(i)$, that the gateway fails to receive. The MLR is computed as \mbox{$[\sum_{i=1}^{n_s}\rho(i)/n_s]^{\tilde r + 1}$}, where $n_s$ is the number of runs. For statistical reliability, $n_s$ is chosen such that the product of the MLR and the total number of simulated transmissions is at least 100.

We simulate the following five schemes that differ in their allocation of redundancy:

\emph{1) No redundancy} (NR):  A transmitted frame contains only the current measurement.

\emph{2) Maximum redundancy} (MR): The number of most recent past measurements per frame is blindly set to $r_{\max}$.

\emph{3) Calculated Redundancy with the Correct Fading model and Uniformly distributed Distance assumption} \mbox{(CR-CF-UD)}: The number of measurements to be included per frame is determined as described in Section~\ref{optimal_r}. The gateway has an accurate estimate of the probability distribution of the fading. However, it does not know the exact distribution of the distances from the gateway to the sensors. Instead, the gateway assumes that the distances are uniformly distributed between its rough estimates of the distances to the nearest and farthest sensors. For our numerical results, these estimates are 44 m and 57 m, respectively. The estimates are not accurate because the coordinates of the sensors are chosen at random for each simulation run.

\emph{4) Calculated Redundancy with the Correct Fading model and Equal-Distance assumption} \mbox{(CR-CF-ED)}: Same as CR-CF-UD except that the gateway assumes that the distances to all sensors are approximately the same, and therefore applies~\eqref{S_I_equal_dist_assumption} and~\eqref{S_F_equal_dist_assumption} to compute $P_i(r)$ and $P_f(r)$, respectively. In the evaluation of~\eqref{S_F_equal_dist_assumption}, the value of $d$ is set to 50.5 m, which is the average of the gateway's distance estimates to the nearest and farthest sensors.

\emph{5) Calculated Redundancy with an Incorrect Fading model and Uniformly distributed Distance assumption} \mbox{(CR-IF-UD)}: Same as CR-CF-UD except that the gateway's estimate of the fading is not an accurate match for the actual fading; it assumes Nakagami with $m \equals 1.5$ rather than $m \equals 1$. In practice, the gateway might update its estimate of the fading as it receives frames from the sensors. However, to model the worst-case scenario, we assume that no such update takes place during the simulation period.

\section{Performance Results}
\label{perf_res}

\begin{figure}
    \subfloat[$P_t = 0.001.$]{\label{FR_vs_N_Pt001}\includegraphics[scale=0.33,bb= -50 0 770 550]{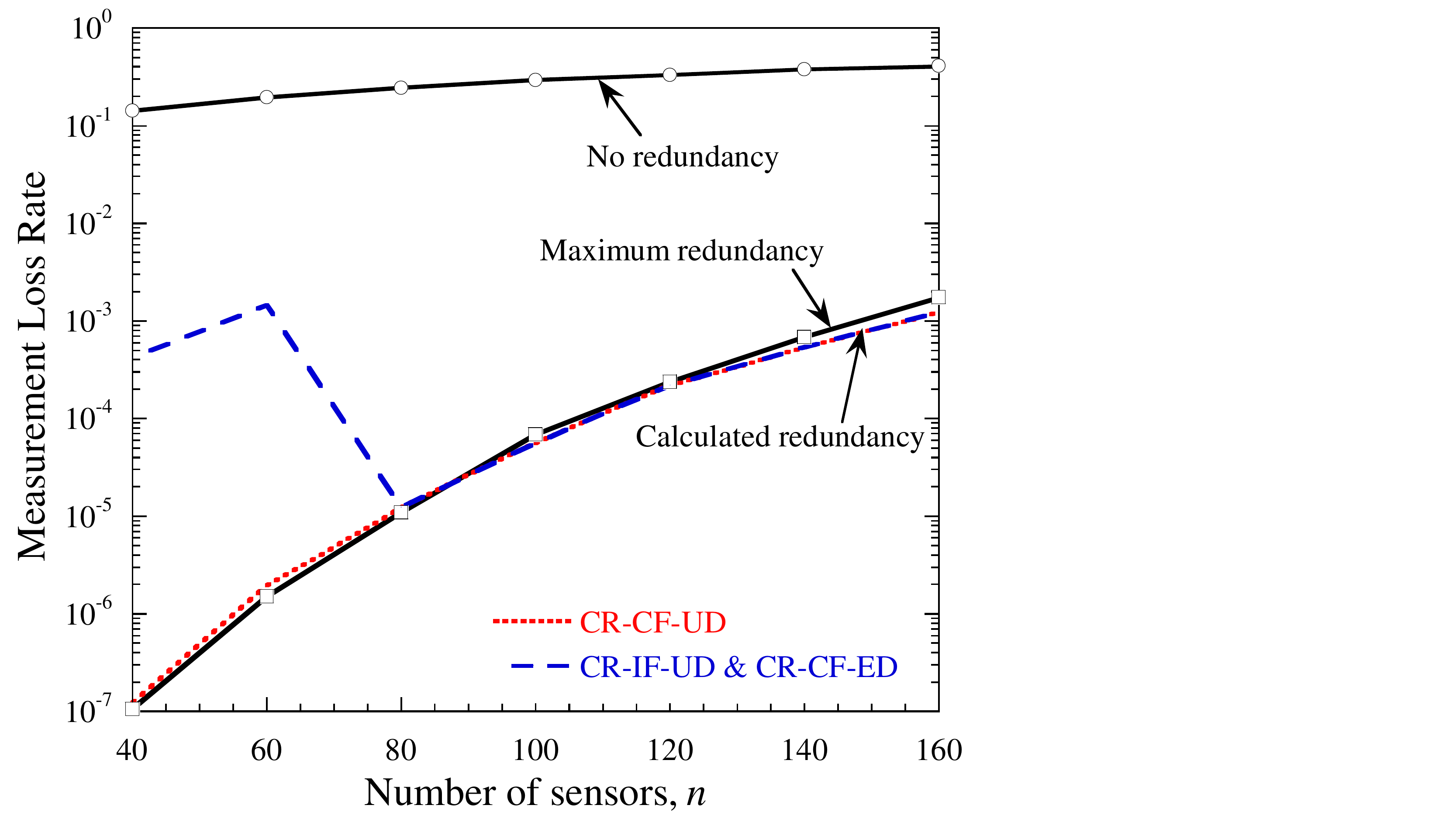}} \\
    \subfloat[$P_t = 0.01.$]{\label{FR_vs_N_Pt01}\includegraphics[scale=0.33,bb= -50 0 770 550]{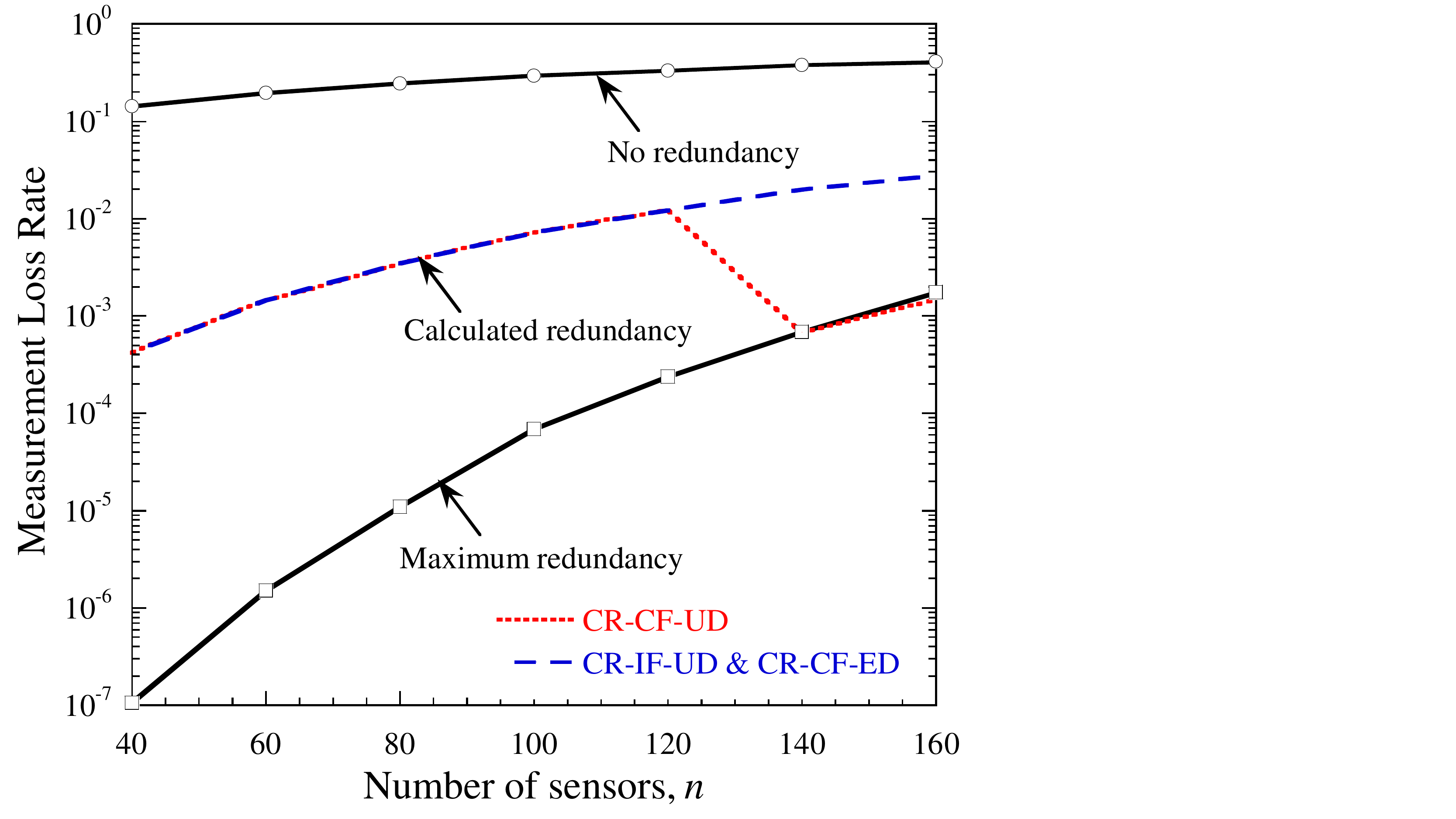}}
    \setlength{\belowcaptionskip}{-15pt}
    \caption{Measurement loss rates for different redundancy schemes.}
    \label{FR_vs_N}
\end{figure}

The MLRs for given target failure probabilities are plotted against the number of sensors $n$ in Fig.~\ref{FR_vs_N}. Results are shown for $n \equals 40$ to $160$ in steps of 20. 
We observe that the MLR of NR is high and increases from 0.14 for $n \equals 40$ to 0.41 for $n \equals 160$. This deterioration is a result of increased interference. Each of the four schemes that employs redundancy greatly outperforms NR. In Fig.~\ref{FR_vs_N_Pt001} ($P_t = 0.001$), \mbox{CR-CF-UD} outperforms NR by up to six orders of magnitude. The performance of \mbox{CR-CF-UD} is very close to that of MR for all $n$. For $n \gteq 140$, the schemes that calculate the redundancy slightly outperform MR. Since MR sends more redundancy compared to others, its frames are longer and experience more collisions. The performance degradation of MR is expected to become progressively worse if $n$ increases beyond 160. This demonstrates that blindly transmitting more redundancy is not always beneficial. Both  \mbox{CR-CF-ED}  and  \mbox{CR-IF-UD} give the same MLR and are inferior to \mbox{CR-CF-UD} at $n \equals 40$ and $n \equals 60$. For other values of $n$, all three schemes that calculate the redundancy provide the same MLR. Even at points where the MLRs of  \mbox{CR-CF-ED} and \mbox{CR-IF-UD} are higher than that of \mbox{CR-CF-UD}, they still achieve MLR close to or lower than the target $P_t$. The MLR of the schemes for $P_t = 0.01$ are compared in Fig.~\ref{FR_vs_N_Pt01}. As expected, an increase in the value of $P_t$ increases the MLR of all three CR schemes. However, the MLR of \mbox{CR-CF-UD} is approximately the same as that of MR for $n = 140$ and $n = 160$.

Table~\ref{r_table} lists the number of redundant measurements $\tilde r$ transmitted by the CR schemes for $P_t = 0.001$. Also shown are the values of $r^*$. The sudden jumps in $\tilde r$ from 3 to 8 at $n \equals 80$ for \mbox{CR-CF-ED} and  \mbox{CR-IF-UD} explain the sharp drops in the curves for these two schemes in Fig.~\ref{FR_vs_N_Pt001}. We observe that, although the three schemes frequently compute different values for $r^*$, the redundancy $\tilde r$ that is actually sent is often the same. As explained in Section~\ref{optimal_r}, this is because our strategy exploits the fact that LoRa frames can have the same duration for more than one payload size. This feature ensures that the strategy is not very sensitive to errors in the estimates of the parameters used in the computation of~$\tilde r$. Therefore, in order to allocate redundancy, the gateway can store a table of $\tilde r$ values computed using the equal-distance approximation for different distances and different numbers of nodes. In a static industrial environment, we do  not expect the probability distribution of the fading to change rapidly over time. Therefore, a previously obtained estimate of the fading (e.g., from channel measurements performed prior to setting up the network) can be used for the computation. The gateway can perform a table lookup to allocate redundancy when there is a change in the network (e.g., when new sensors are installed). A re-calculation of $\tilde r$ is required only if the MLR exceeds certain application-dependent threshold.

\begin{table}
\scriptsize
\centering
\caption{Values of $r^*$ and $\tilde r$ for $p_t \equals 0.001$.}
\begin{tabular}{|>{\centering}p{5mm}|>{\centering}p{5mm}|>{\centering}p{5mm}|>{\centering}p{5mm}|>{\centering}p{5mm}|>{\centering}p{5mm}|>{\hspace{2mm}}p{5mm}|}
\hline
\multirow{2}{*}{$n$} & \multicolumn{2}{c|}{CR-CF-UD} & %
    \multicolumn{2}{c|}{CR-IF-UD} & \multicolumn{2}{c|}{CR-CF-ED}\\
\cline{2-7}
  & $r^*$ & $\tilde r$ & $r^*$ & $\tilde r$ & $r^*$ & $\tilde r$\\
\hline
 40 & 4 & 8 & 3 & 3 & 3 & 3\\
\hline
 60 & 4 & 8 & 3 & 3 & 3 & 3\\
\hline
 80 & 5 & 8 & 4 & 8 & 4 & 8 \\
\hline
100 & 5 & 8 & 5 & 8 & 4 & 8 \\
\hline
 120 & 6 & 8 & 5 & 8 & 4 & 8 \\
\hline
 140 & 6 & 8 & 6 & 8 & 5 & 8\\
 \hline
 160 & 7 & 8 & 6 & 8 & 5 & 8\\
 \hline
\end{tabular}
\label{r_table}
\end{table}
\renewcommand{\baselinestretch}{0.9}
\small\normalsize

\begin{figure}
\centering
\includegraphics[scale=0.33,bb= 0 0 770 550]{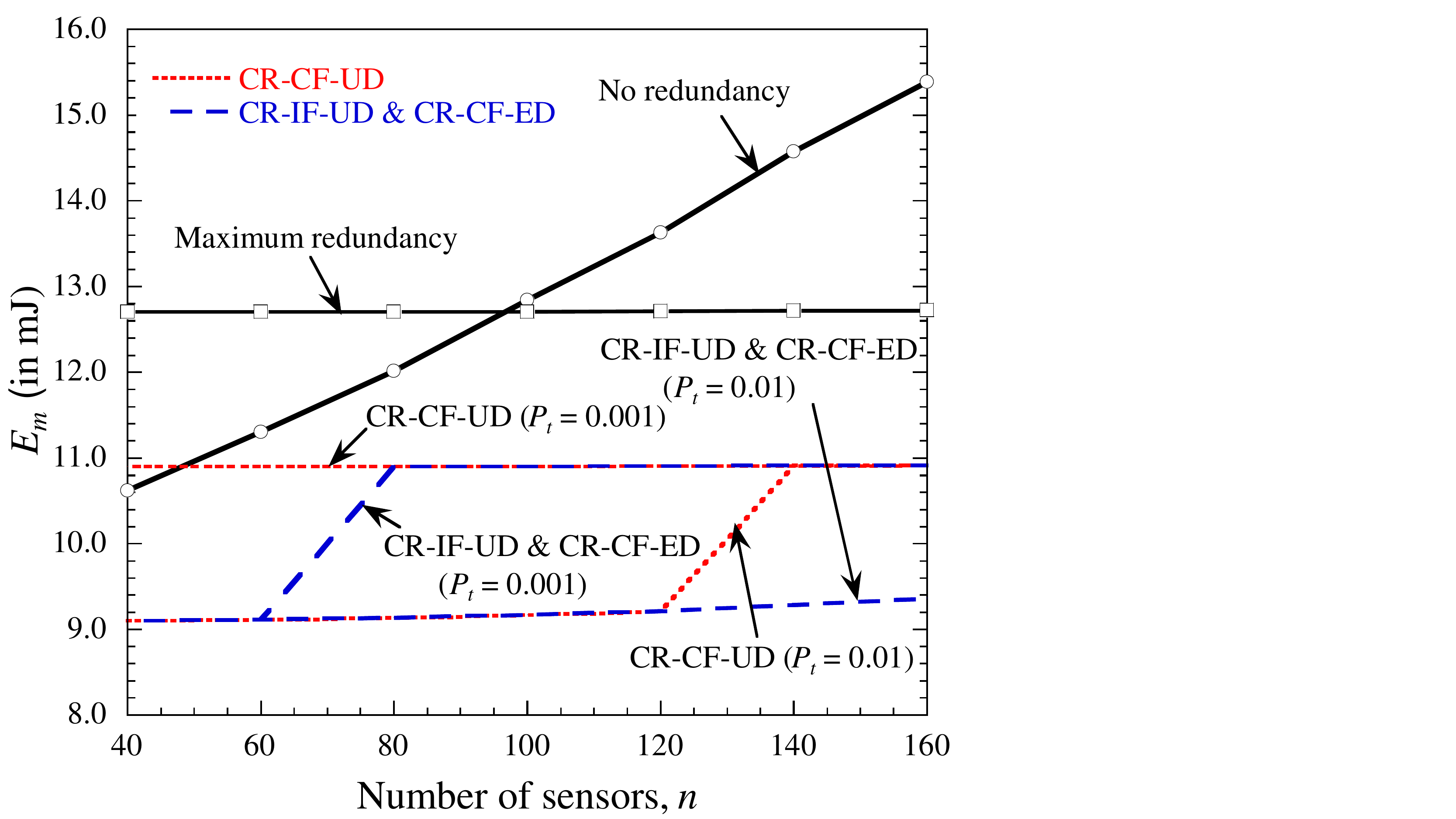}
\setlength{\belowcaptionskip}{-12pt}
\caption{Energy expenditure per delivered measurement for different redundancy schemes.}
\label{EDM_vs_N}
\end{figure}

Fig.~\ref{EDM_vs_N} shows the energy expenditure of all schemes. The $E_m$ of NR increases sharply with $n$ because of increased interference. The $E_m$ of MR is consistently higher (by up to $40\%$) than every scheme that calculates the redundancy. For systems with larger $r_{\max}$, the $E_m$ of MR would be even poorer with hardly any decrease in the MLR; in fact, as pointed out earlier, even the MLR can increase for large $n$. For the schemes that calculate the redundancy, a smaller $P_t$ yields a higher $E_m$, since more redundancy is sent to achieve a lower failure probability. For applications with higher tolerance for undelivered measurements, larger values of $P_t$ can be employed to achieve smaller $E_m$. Thus, our approach provides a mechanism to flexibly allocate redundancy according to the needs of the application.

\renewcommand{\baselinestretch}{0.9}
\small\normalsize

In addition to the results reported above, we simulated other scenarios in which the sensors are distributed over a larger area (100 sensors randomly placed over a \mbox{30 m $\times$ 30 m} region). Results of these simulations show that our procedure for redundancy allocation consistently provides lower MLR than NR and lower $E_m$ than both NR and MR. \mbox{CR-CF-ED}, in which the gateway assumes all sensors to be approximately at the same distance, performs close to \mbox{CR-CF-UD} despite the larger area. The reason is as described earlier: Even when the equal-distance assumption of \mbox{CR-CF-ED} yields a different $r^*$ than the uniform-distance assumption, the value of $\tilde r$ might still be the same, thereby resulting in similar MLR and $E_m$.

We finally note that although we illustrate our approach with the target failure probability $P_t$ as the design criterion, one can employ the same analytical framework to optimize other metrics. For example, if it is desired to achieve a certain $E_m$, we can determine the expected $E_m$ by computing \mbox{$\overline{E}_m(r) \equals \cE(r)/(1-P_f(r))$}, where $P_f(r)$ is defined in Section~\ref{reliability_analysis}, and $\cE(r)$ is the energy of a frame that contains one current and $r$ past measurements. Next, the desired $\tilde r$ can be obtained by replacing $P_f(r)$ with $\overline{E}_m(r)$, and $P_t$ with the target $E_m$ throughout Section~\ref{optimal_r}.

\section{Conclusions and Outlook}
\label{conclusion}
Repetition redundancy can be used to improve the reliability of LoRa sensor networks. We devised a method to determine the redundancy that must be transmitted to achieve a target performance. Numerical results show that the transmission of redundancy substantially reduces the fraction of sensor measurements that fail to reach the gateway. Our method for redundancy allocation, which takes into account the effects of fading and interference, incurs lower energy expenditure per delivered measurement as compared to a scheme that blindly allocates the redundancy. Good performance is achieved even without accurate knowledge of the fading and node locations.

As future work, we plan to explore the possibility of extending our strategy to systems that generate the redundancy via application-layer coding. Derivation of closed-form approximations for the integrals that must be evaluated to calculate the redundancy is also of interest.



\renewcommand{\baselinestretch}{0.85}

\end{document}